\documentstyle[aps,prb]{revtex}
	
\begin{document}
\input epsf.sty
\twocolumn[\hsize\textwidth\columnwidth\hsize\csname %
@twocolumnfalse\endcsname]
\draft
\widetext

%
%
%

\title{X-ray scattering study of two length scales in the critical fluctuations of CuGeO$_3$}

\author{Y.J. Wang, Y.-J. Kim, R.J. Christianson, S.C. LaMarra, F.C.Chou, and R.J. Birgeneau}
\address{Department of Physics, Massachusetts Institute of Technology,
Cambridge, MA 02139}

\date{\today}
\maketitle

\begin{abstract}
  The critical fluctuations of CuGeO$_3$ have been measured by synchrotron 
x-ray scattering, and two length scales are clearly observed. The ratio between
 the two length scales is found to be significantly different along the $a$ 
axis, with the $a$ axis along the surface normal direction. We believe that 
such a 
directional preference is a clear sign that surface random strains, especially
 those caused by dislocations, are the origin of the long length scale 
fluctuations.      

\end{abstract}

\vspace{0.5in}


\section{Introduction}

 High resolution x-ray and neutron scattering studies of the critical 
fluctuations
 associated with structural and magnetic phase transitions typically 
reveal ``two 
length scales'', that is, two distinctive scattering lineshapes superimposed upon each other in the critical scattering profile\cite{Cowley}. Since the existence of 
the second length scale seems to contradict the fundamental assumption of 
modern critical phenomena theory that there exists only one characteristic 
length in the critical fluctuations, extensive experimental
 and theoretical efforts have been devoted to elucidating the exact origin of this phenomenon. However, in spite of a significant amount of work 
dedicated to this problem, 
 a consensus still has not been reached. Presently, there exist two main approaches\cite{Cowley}:(1) models based on intrinsic near-surface effects and (2) explanations 
involving near-surface 
random 
defects.
The accumulating experimental evidence seems to favor the randomness interpretation although there is still no  definitive experiment to pinpoint the exact 
origin of the second length scale fluctuations.

In the present paper, we present a high-resolution synchrotron x-ray 
scattering study of the critical fluctuations associated with the spin-Peierls 
structual phase transition in CuGeO$_3$. Not only do we clearly observe two  
lineshapes in the critical scattering profile, but we also observe a dramatic 
change of the 
anisotropy ratio of the correlation length divergence along the three primary 
crystal axes. The existence of the modified anisotropy ratio provides 
substantial evidence that near-surface dislocations are the origin of the 
second length scale fluctuations.

Our paper is organized as follows: In Section II we provide details of the sample preparation and experimental measurements. In Section III we present our experimental results. A discussion of the results and conclusions is given in Section IV.

\section{Experimental Procedures}

  The experiment was carried out at MIT-IBM beamline X20A at the 
National Synchrotron Light Source. The x-ray beam was focused by a mirror,
monochromatized by a pair of Ge (111) crystals, scattered from the sample,
and analyzed by a Si (111) analyzer. The x-ray energy was 8.5 keV. High
quality pure CuGeO$_3$ and Cu$_{0.99}$Zn$_{0.01}$GeO$_3$ single crystals 
grown by the travelling 
solvent floating zone
 method were used. Carefully cleaved samples
were placed inside a Be can filled with helium heat-exchange gas and
mounted on the cold finger of a 4K closed cycle cryostat.   
The experiment was carried out around
the (1.5, 1, 1.5) SP dimerization peak position with the
(H K H) zone in the scattering plane. 

\section{Experimental Results} 

 Pretransitional lattice fluctuations along the H, K and L directions have been
 measured in pure CuGeO$_3$ by x-ray\cite{Schoeffel,Harris} and
 neutron scattering\cite{Hirota1}. All of the experiments have shown rapid and
 anisotropic broadening of the scattering peaks when the sample was heated 
across
 T$_{SP}$, which was clear evidence of anisotropy in the magnetic interaction\cite{Schoeffel}. Close to T$_{SP}$, however, Schoeffel {\it et al.}\cite{Schoeffel} observed a crossover temperature T$_{CO}$ where the ratio 
of the correlation lengths along the three crystal axis directions appeared to change abruptly:
 $\xi_{c}/\xi_{a}\sim 4$ and $\xi_c/\xi_b\sim1$ below $T_{CO}$ and $\xi_c/\xi_b\sim1.6$ 
above. This was used as evidence of a crossover to a 2D lattice
fluctuation regime above  $T_{SP}$. Later, both experimental and 
theoretical efforts were devoted to elucidating the exact nature of the 2D 
crossover\cite{Lorenzo,Birgeneau}.  Harris {\it et al.}\cite{Harris}, on the other
 hand, studied the critical behavior in the immediate vicinity above T$_{SP}$ and 
reported a different anisotropy ratio. However, the critical 
fluctuations reported by Harris {\it et al.}\cite{Harris} have length scales which are about an order 
of magnitude larger than those reported by Schoeffel {\it et al.}\cite{Schoeffel}. 
The discrepancies in these two experiments demonstrate that one must treat the
 data near the transition more cautiously. In extracting the correlation length just above T$_{SP}$, it is necessary to take into account explicitly that 
there exist two distinct scattering length scales. Distinguishing and separating their 
individual contributions to the total cross-section will be of primary importance. This comprises a principal motivation of this experiment.

   To reconcile the results of previous critical scattering studies of CuGeO$_3$\cite{Schoeffel,Harris} and to obtain some insight into the physical origin of
 the
 second length scale, we carefully studied the pretransitional critical 
behavior just above 
T$_{SP}$. Though the exact origin of the long length scale fluctuations has not
 been 
determined, it has long been speculated that they originate from random surface stresses caused
 by defects\cite{Cowley,Harris}. Hence, we
 prepared our samples by cleaving them several times until no observable cracks could be seen by visual inspection. In doing so, we took advantage of the fact that
 CuGeO$_3$ crystals are inclined to self-cleave along the $a$ crystal plane.
 Thus,
 no additional 
grinding or polishing process is necessary to achieve a visually smooth mirror 
surface. To put the two previous seemingly conflicting experiments together,
 we need to have information on both length scales in the same sample. 
Fortunately, this is
 exactly what we have observed in our experiment. In Fig. 1 we show the critical scattering profiles 
along the H, K, and L directions at T$_{SP}+0.1$K and T$_{SP}+0.3$K for undoped 
CuGeO$_3$. 
At T$_{SP}+0.1$K, there clearly exist two distinct scattering profiles along all three directions, with a sharp central peak superimposed upon a broader peak.
This corresponds to archetypal two-length scale behavior. However, a 
closer examination of the data reveals that even though there are 
clearly two features along all three directions, the central peak along the H 
direction
 is much sharper in comparison to the broad one than is observed along the 
other two 
directions. In other words, the ratio of the correlation lengths for these two
 length scales are significantly different along one of the three crystal axis 
directions.

 In Fig. 2 , we show the inverse correlation lengths of the broad component 
as functions of temperature along the H, K, and L directions. Several features
 can be 
recognized
 immediately. First, the correlation length diverges rapidly as the 
temperature approaches T$_{SP}$ from above, which demonstrates that the SP 
transition in our CuGeO$_3$ crystal is a  well defined second-order phase transition. Second, the correlation length 
also diverges anisotropically along the three crystal axes. In the temperature 
range $T_{SP}<T<T_{SP}+0.4K$, 
the anisotropy ratio remains  $\xi_c/\xi_a\sim 5$ and $\xi_c/\xi_b\sim 3$, 
which is consistent with the high temperature data taken both by x-ray and 
neutron scattering\cite{Schoeffel,Hirota1}. Thus we do not observe any evidence for the 
presumed 2D 
crossover \cite{Schoeffel}, in which there should exist a 
dramatic change in the anisotropy ratio about 1K above T$_{SP}$. Specifically,
in Ref\cite{Schoeffel} it is argued that below this 
crossover temperature, the correlation length along the $b$-axis direction 
equals the correlation length along $c$-axis direction. Our 
results clearly demonstrate that the correlation length anisotropy ratio 
remains unchanged from high temperatures to very near T$_{SP}$, that is, 
there is no evidence for any crossover.

Fig. 3 shows the inverse correlation length of the sharp component as a function 
of temperature. One of the salient features is that, 
although the correlation length of the sharp component diverges in a manner similar to that of the broader 
component, the anisotropy ratio of the 
correlation lengths along the three axes directions is modified to $\xi_c/\xi_a\sim 1.5$
 and $\xi_c/\xi_b\sim 4.4$. This is reminiscent of the high resolution results 
reported by 
Harris {\it et al.}\cite{Harris}. Instead of the relationship  $\xi_c>\xi_b>\xi_a$, the large length scale fluctuations
 exhibit the hierarchy $\xi_c>\xi_a>\xi_b$. The change of the order of the correlation 
lengths is informative, since there are not  many physical mechanisms which 
could induce such a directional preference. The confirmation of the change of 
the order of the correlation lengths in our experiment proves that this is a 
general 
phenomenon 
instead of an irreproducible singular case. Furthermore, if we directly compare the magnitude of the two length scales along $a$, $b$ and $c$ crystal axes,  ratios of $28:6:8$ would result, with the maximum along the $a$ axis and similar values along the $b$- and $c$-axes. The other feature worth mentioning is the relative 
importance of the second length scale in both studies. In Harris {\it et. al.}
 \cite{Harris}'s case, only the long length scale fluctuations were clearly  
observable over the temperature range studied. On the other hand, in our experiments, the 
fluctuations associated with  both length scales
 are clearly observable, which proves that the relative amplitude of the second
 length scale fluctuations is
sample dependent.

  Over the last several years, we and others have carried out detailed studies of the effects of dopants on the CuGeO$_3$ magnetic and structural phase transitions with a focus on the overall phase diagram\cite{Wang1,Masuda,Wang3}. Such studies can be regarded as  a systematic exploration of the effects of
 point defects
 on the CuGeO$_3$ structural phase transition. Thus, as a byproduct of our Cu$_{1-x}$(Zn,Mg)$_x$GeO$_3$ phase diagram studies, we also are able to test the hypothesis that the long length scale fluctuations are caused by point defects\cite{Trenkler}.
  
 Fig. 4 shows the inverse correlation lengths along the three crystal axes as 
functions of temperature for 1\% Zn-doped CuGeO$_3$. The dramatic effects of 
the Cu ion dilution  on the phase transition are apparent: the transition temperature has been suppressed by more than
 1K upon only 1\% Zn doping, and the critical exponent associated with the correlation length appears to be different from that of the undoped sample. We are 
uncertain currently whether this apparently different critical behavior is intrinsic or merely due to a trivial concentration gradient effect. Further experiments are needed to clarify this issue. 
However, the ratios of the inverse correlation lengths are $\xi_{c}:\xi_{b}:\xi_{a} = 5.9:2.0:1$, which are essentially identical to those of the undoped samples. This consistency of 
the anisotropy ratios between the doped and undoped sample naturally excludes models for the second length scale based on point defects.

\section{Discussion}

Before we present our interpretion of our experimental observations, we first briefly summarize the results of previous experimental and 
theoretical studies on the two length scale phenomenon. Most high resolution 
critical scattering studies
of both structural and magnetic phase transitions reveal two length scales\cite{Cowley}. Further, an elegant neutron scattering study by 
Shirane and coworkers reveals that the long 
length scale fluctuations are located in the ``skin'' of the sample\cite{Hirota2}. A subsequent
 study on the same single crystals by transmission electron microsopy[TEM]\cite{Wang2} finds that the density of dislocations has a steep increase within
 a few microns of the sample 
surface, which coincides with the onset of the long length scale. Based on the
 spatial coexistence of the second length scale and dislocations, the authors of Ref\cite{Wang2} conclude that the second length scale originates 
from dislocations, albeit in an indirect way. As more and more experimental evidence turns up, a gradual consensus is emerging that the origin of the second 
length scale fluctuations is the 
random strain fields caused by defects in the sample skins\cite{Cowley}. 
However, an intrinsic effect explanation can not be excluded\cite{Cowley}. 
Moreover, even if the idea that the second length scale originates from defects is taken for granted, there exists additional complexity because the defects
 can either be point defects or line defects such as dislocations. A recent 
study suggests that point defects are responsible for the occurrence of the second length scale\cite{Trenkler}. The dislocation theory, on the other hand, 
has been
 less favored. One of the key objections used against it is the lack
 of directional preference\cite{Cowley} in all the previous studies, that is, 
dislocations are line defects and they should inevitably favor particular directions.            
From the results of our study, we believe that CuGeO$_3$ serves as a model system to study the origin of the second length scale and provides strong evidence that dislocation defects are responsible for the occurence of the second length scale fluctuations. 

 Using dislocation theory, in the following, we explain our experimental 
observations by a phenomenological model. One of the marked differences between our results and those reported by 
Harris\cite{Harris} is the relative importance of the long length scale fluctuations. This 
can easily be explained, since the density and spatial distribution of 
dislocations naturally depend on sample preparation and surface processing 
such as chemical etching, so they would unavoidably vary from sample to sample.
 The most determinant piece of information to support a dislocation model is the occurrence of a
directional preference. The experimental results on the Zn-doped sample provide additional support by demonstrating the irrelevancy of point defects. To 
understand qualitatively the experimentally observed direction preference, we 
refer to the theoretical work by Altarelli {\it et al.}\cite{Altarelli}, in 
which the effect of dislocations has been treated on a qualitative level. 
As discussed by Altarelli {\it et al.}\cite{Altarelli}, in real crystals, surface treatment always induces slipping parallel to the surface. These defects are
 edge dislocations parallel to the sample surface but randomly oriented in the
 plane. They induce anisotropic stress fields in the surrounding crystal since they are line
 defects by nature.
 The stress 
field produced by dislocations can be well modeled by dipole fields with the 
maximum in the plane perpendicular to the dipole, which is the Burgers vector direction in our case. The whole problem can then be mapped into that of a group 
of randomly oriented dipoles lying in a plane. The stress field 
can lower the free energy of the structually ordered phase, thus increasing 
the phase transition temperature in the stressed region. This is used to account for the emergence of the second length scale\cite{Altarelli,Wang2}.

  Using this theory, the different ratio between the two length scales can  be 
qualitatively explained. We recall that CuGeO$_3$ crystals naturally cleave in
 the $a$ plane.The difference in magnitude of the two length scales is most 
prominent along the $a$-axis because the fluctuation amplitude is presumed to 
be proportional to the average stress field. The 
random orientation of the dipoles in the surface would results in an isotropic 
stress field distribution in the plane. However, the maximum average stress 
field would
 be produced along the surface normal direction due to the dipole nature. We believe that the stress field is responsible for the creation of 
pretransitional ordered domain structures\cite{Wang2}. These domains order at 
a higher temperature than the bulk and  have an anisotropic structure owing to the anisotropy of the stress field. This 
can naturally explain the unusual sharp feature of the critical scattering 
along $a$-axis and also why the ratio of the two length scales remains relatively
 unmodified in the other two directions.

We find a tiny difference in the ratio 
along the $b$ and $c$ directions, in agreement with Harris {\it et al.}\cite{Harris}. We speculate that this subtle anisotropy 
originates in a slight anisotropic distribution of the dislocations in the plane. Indeed, a closer inspection of a naturally 
cleaved CuGeO$_3$ sample surface reveals that the apparently smooth 
surface
 is actually composed of  some stripes running along the $c$-axis direction. These are 
most likely formed during crystal growth. When the single crystals are grown using 
the floating zone method, the seed rod is oriented with the easy growth direction 
coinciding with the travelling zone direction. Stripes are then naturally formed
 along the direction of the crystal growth, which is $c$-axis. From the theory of dislocations, 
structural line defects are preferentially created along the same direction. 
These defects are normally edge dislocations with the Burgers' vector 
perpendicular to the dislocation line and lying in the slip plane, the $b$-axis
 direction in our 
case. Hence , this could create  a tiny preference for the dipoles to lie in the
 $b$ direction: a resulting minimum ratio along the $b$-axis is expected.

 We would further comment that even though dislocation theory offers a 
satisfactory heuristic explanation of our critical scattering results, many open questions still exist, for example, why is there a similar ratio between these two length scales in many different physical systems and why there does there exist a clear phase transition for the long length scale fluctuations despite the fact we are assuming a spread of transition temperatures. More theoretical and experimental work is needed to address these issues.

In conclusion, we have studied the critical fluctuations in pure and Zn doped 
CuGeO$_3$. Two length 
scales have been observed with different anisotropy ratios for the correlation lengths along the three crystal axes. The maximum of the magnitude of the two length scales is found to be along the $a$-axis direction, which is the surface normal of the crystal. We argue that dislocation theory would serve as the 
best explanation of the origin of the second length scale fluctuations.

\section{Acknowledgment}
  We thank G. Shirane for insightful comments. This work was
supported by the NSF-LTP Program under Grant No. DMR97-04532.


\begin{figure}
\centerline{\epsfxsize=3.2in\epsfbox{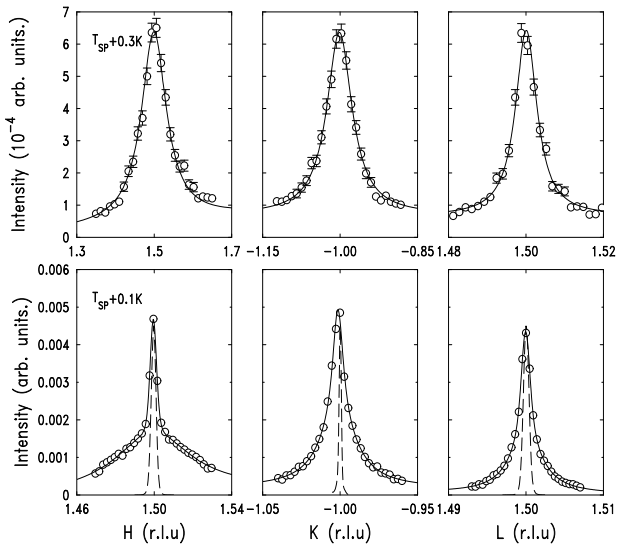}}
\vskip 5mm
\caption{Representative critical scattering scans at the superlattice peak (1.5 -1 1.5) for undoped CuGeO$_3$, the dashed lines represent the instrumental resolution function, the solid lines are fits to the data. In the bottom panel, in the close vicinity of T$_{SP}$, a
 sum of Lorentzian plus Lorentzian squared lineshape is used. In the upper 
panel, much higher than T$_{SP}$, a single Lorentzian lineshape is used, the fits are the results of the 
convolution with the resolution function.}
\label{Fig.1}
\end{figure}

\begin{figure}
\centerline{\epsfxsize=3.2in\epsfbox{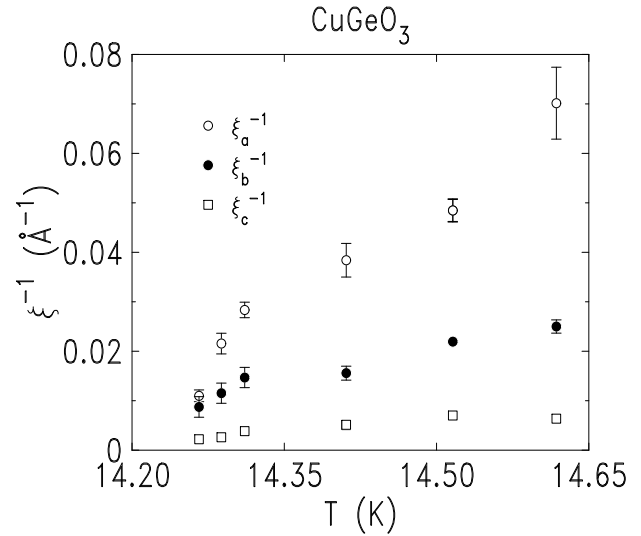}}
\vskip 5mm
\caption{Inverse correlation length along the H, K, and L direction of the 
thermal critical scattering (short length scale feature) for undoped CuGeO$_3$ as functions of temperature.}
\label{Fig.2}
\end{figure}

\begin{figure}
\centerline{\epsfxsize=3.2in\epsfbox{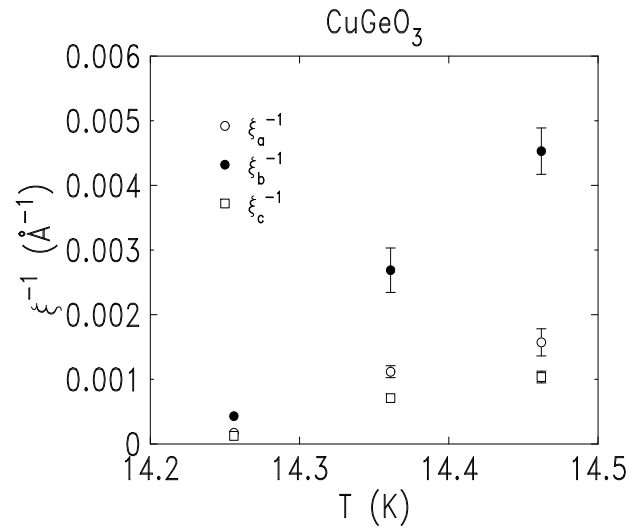}}
\vskip 5mm
\caption{Inverse of the correlation length associated with the long length 
scale fluctuation along the H, K, and L directions as functions of temperature.}
\label{Fig.3}
\end{figure}

\begin{figure}
\centerline{\epsfxsize=3.2in\epsfbox{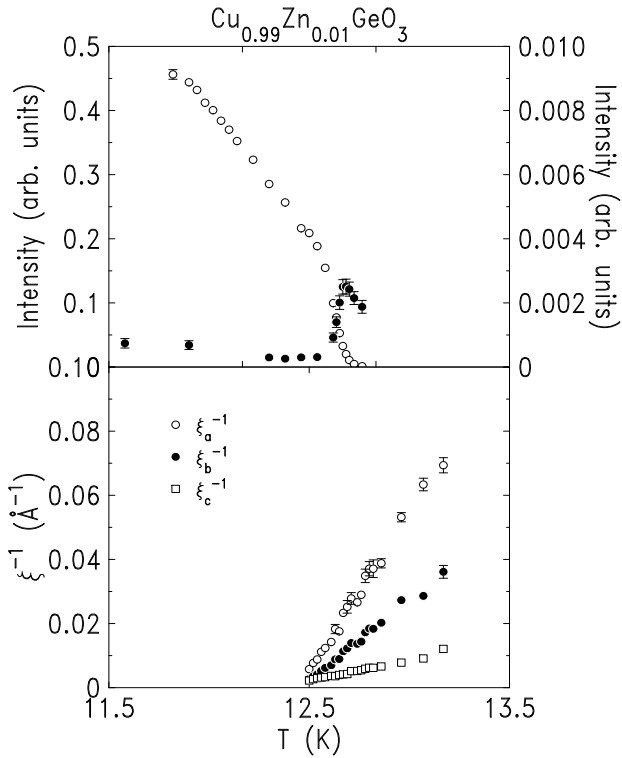}}
\vskip 5mm
\caption{(a) peak intensity of the superlattice peak and critical fluctuation 
intensity at the wing as functions of temperature for x=0.01 Zn doped CuGeO$_3$.
(b) Inverse correlation length along H, K, and  L directions of the thermal 
critical scattering for x=0.01 Zn doped CuGeO$_3$ as functions of temperature.}
\label{Fig.4}
\end{figure}


\end{document}